\documentclass[useAMS]{mn2e}
\usepackage{graphicx,epsfig}

\def\etal{{\it et~al. }} 

\title{On fitting planetary systems in counter-revolving configurations} 

\author[J. Gayon-Markt \& E. Bois]
{Julie Gayon-Markt$^{(2),(1)}$\thanks{E-mail:julie.gayon@oca.eu} 
\& Eric Bois$^{(1)}$\\
$^{(1)}$University of Nice Sophia-Antipolis, CNRS, Observatoire de la C\^ote 
d'Azur, B.P. 4229, F-06304 Nice Cedex 4, France\\
$^{(2)}$NASA Ames Research Center, Space Science and Astrobiology Division, 
MS 245-3, Moffett Field, CA 94035, USA
} 

\begin{document}

\date{Draft version \today}

\maketitle

\begin{abstract}
In Gayon \& Bois (2008) and Gayon \etal (2009), (i) we studied the
theoretical feasibility and efficiency of retrograde mean motion resonances
(i.e. two planets are both in orbital resonance and in counter-revolving
configuration), (ii) we showed that retrograde resonances can generate
interesting mechanisms of stability, and (iii) we obtained a dynamical fit
involving a counter-revolving configuration that is consistent with the 
observations of the HD\thinspace73526 planetary system. 
In the present paper, we present and analyze data reductions assuming
counter-revolving configurations for eight compact multi-planetary systems
detected through the radial velocity method.
In each case, we select the best fit leading to a dynamically stable
solution.  
The resulting data reductions obtained in $rms$ and $\sqrt{\chi_\nu^2}$ values 
for counter-revolving configurations are of the same order, and sometimes
slightly better, than for prograde configurations.
In the end, these fits tend to show that, over the eight studied
multi-planetary systems, six of them could be regulated by a mechanism
involving a counter-revolving configuration.

\end{abstract}

\begin{keywords}
planetary systems - techniques: radial velocities - 
stars: individual (HD37124, HD69830, HD73526, HD108874, HD128311, HD155358, 
HD160691, HD202206) 
\end{keywords}

\section{Introduction}
The orbital element determination of extrasolar planets from radial velocity
measurements is relatively complex. As mentioned in Beaug\'e \etal (2008),  
the equations relating observations to orbital elements (and minimal planetary
masses) are highly non-linear and generate different local minima in 
the parameter space, and consequently, different possible observational fits.
Moreover, in order to correctly determine orbital elements, the ratio between
the $N$ number of observations and the $M$ number of free parameters must be
relatively high. But generally, the duration of observations is only of the 
order of $2$ or $3$ times the orbital period of the outer planet of a system. 

Owing to the necessity of observing systems over a large number of times the 
outer planet period (in order to determine orbital elements with a convenient 
precision), the assurance of a correct determination of orbital elements is
not necessarily guaranteed. The real dynamics of multi-planetary systems found 
until now is consequently difficult to point out. At this time, the orbital
elements of only one multi-planetary system prove to be acquired~: the very
compact Gliese 876 system. 
Known since 1998 (Marcy \etal 1998, Delfosse \etal 1998,
Marcy \etal 2001, Rivera \etal 2005), a large series of observations has
allowed to gather a sufficient number of radial velocity measurements to 
determine with a good precision the orbital elements of the
Gliese 876 main planets. Such a determination has continually been improved 
since 1998. While the two major planets are revolving around
their host star in about $30$ days and $60$ days, observations have been
performed for $7$ years. The $N/P$ ratio between the number of
observations ($N$) and the orbital period ($P$) of planets is then
particularly high and permits a good precision of the Gliese 876 system fit. 
Unfortunately, the whole of other detected
multi-planetary systems does not present such a high $N/P$ ratio. 
The orbital determination of all the other systems is not still completely 
acquired.
In the present paper, we propose to carry out new observational fits
for specific configurations of several compact multi-planetary systems.
For eight compact planetary systems (HD\thinspace37124, 
HD\thinspace69830, HD\thinspace73526, HD\thinspace108874, HD\thinspace128311, 
HD\thinspace155358, HD\thinspace160691, and HD\thinspace202206), we indeed 
assume that one planet of each system moves in retrograde direction on its
orbit, while other planets have a prograde motion around the host star.

\begin{table*}
\begin{center}
\caption{\label{tab_obs}New data reductions obtained for counter-revolving
  configurations from the following planetary systems~:
HD\thinspace37124 ($M_*=0.78\, M_\odot$),
HD\thinspace69830 ($M_*=0.86\, M_\odot$),
HD\thinspace73526 ($M_*=1.08\, M_\odot$), 
HD\thinspace108874 ($M_*=0.99\, M_\odot$),
HD\thinspace128311 ($M_*=0.84\, M_\odot$), 
HD\thinspace155358 ($M_*=0.87\, M_\odot$),
HD\thinspace160691 ($M_*=1.15\, M_\odot$), and 
HD\thinspace202206 ($M_*=1.15\, M_\odot$).
Non-zero stellar jitters were used for several planetary
systems~: HD\thinspace37124 ($3.2\,m.s^{-1}$), HD\thinspace108874 
($3.9\,m.s^{-1}$), HD\thinspace128311 ($8.9\,m.s^{-1}$), HD\thinspace155358 
($5.0\,m.s^{-1}$).}

   \begin{tabular}{llccccccc}
   \hline 

    System& &
    $\begin{array}{c}
       m_{P}\\
       (M_{Jup})
    \end{array}$&
    $\begin{array}{c}
       P\\
       (\textrm{days})
    \end{array}$&
    $\begin{array}{c}
       a\\
       (\textrm{AU})
    \end{array}$&
    $\begin{array}{c}
       e\\
       \textrm{}
    \end{array}$&
    $\begin{array}{c}
       i\\
       (\textrm{deg})
    \end{array}$&    
    $\begin{array}{c}
       \omega\\
       (\textrm{deg})
    \end{array}$&    
    $\begin{array}{c}
       M\\
       (\textrm{deg})
    \end{array}$
\tabularnewline
\hline \hline

     HD\thinspace37124 & $\begin{array}{c} b\\ c\\ d\end{array}$&
     $\begin{array}{c}
        0.2059\\
        0.5894\\
        0.7575
     \end{array}$&
     $\begin{array}{c}
       29.377\\
       155.332\\
       841.881
     \end{array}$&
     $\begin{array}{c}
       0.1715\\
       0.5207\\
       1.6067
     \end{array}$&
     $\begin{array}{c}
       0.5155\\
       0.1184\\
       0.0628
     \end{array}$&
     $\begin{array}{c}
        0.0\\
        180.0\\
        0.0
     \end{array}$&
     $\begin{array}{c}
       250.725\\
       266.521\\
       49.524
     \end{array}$&
     $\begin{array}{c}
       64.772\\
       334.402\\
       325.545
     \end{array}$
\tabularnewline
\hline

     HD\thinspace69830 & $\begin{array}{c} b\\ c\\ d\end{array}$&
     $\begin{array}{c}
        0.0318\\
        0.0375\\
        0.0583
     \end{array}$&
     $\begin{array}{c}
       8.666\\
       31.563\\
       197.992
     \end{array}$&
     $\begin{array}{c}
       0.0785\\
       0.1859\\
       0.6322
     \end{array}$&
     $\begin{array}{c}
       0.0955\\
       0.1278\\
       0.0110
     \end{array}$&
     $\begin{array}{c}
        0.0\\
        0.0\\
        180.0
     \end{array}$&
     $\begin{array}{c}
       339.102\\
       216.406\\
       90.368
     \end{array}$&
     $\begin{array}{c}
       264.170\\
       81.289\\
       276.112
     \end{array}$
\tabularnewline
\hline

     HD\thinspace73526 & $\begin{array}{c} b\\ c\end{array}$&
     $\begin{array}{c}
        2.4921\\
        2.5919
     \end{array}$&
     $\begin{array}{c}
        187.935\\
        379.795
     \end{array}$&
     $\begin{array}{c}
        0.6593\\
        1.0538
     \end{array}$&
     $\begin{array}{c}
        0.2401\\
        0.2048
     \end{array}$&
     $\begin{array}{c}
        0.0\\
        180.0
     \end{array}$&
     $\begin{array}{c}
        184.569\\
        58.545
     \end{array}$&
     $\begin{array}{c}
        97.297\\
        221.361
     \end{array}$
\tabularnewline
\hline

     HD\thinspace108874 & $\begin{array}{c} b\\ c\end{array}$&
     $\begin{array}{c}
        1.2141\\
        0.8979
     \end{array}$&
     $\begin{array}{c}
        395.452\\
        1588.626
     \end{array}$&
     $\begin{array}{c}
        0.9953\\
        2.5149
     \end{array}$&
     $\begin{array}{c}
        0.0580\\
        0.2497
     \end{array}$&
     $\begin{array}{c}
        0.0\\
        180.0
     \end{array}$&
     $\begin{array}{c}
        92.572\\
        17.102
     \end{array}$&
     $\begin{array}{c}
        355.512\\
        27.604
     \end{array}$
\tabularnewline
\hline

     HD\thinspace128311 & $\begin{array}{c} b\\ c\end{array}$&
     $\begin{array}{c}
        1.5571\\
        3.2205
     \end{array}$&
     $\begin{array}{c}
        453.626\\
        941.213
     \end{array}$&
     $\begin{array}{c}
        1.0908\\
        1.7756
     \end{array}$&
     $\begin{array}{c}
        0.3550\\
        0.1485
     \end{array}$&
     $\begin{array}{c}
        180.0\\
        0.0
     \end{array}$&
     $\begin{array}{c}
        278.933\\
        49.517
     \end{array}$&
     $\begin{array}{c}
        168.259\\
        235.211
     \end{array}$
\tabularnewline
\hline

     HD\thinspace155358 & $\begin{array}{c} b\\ c\end{array}$&
     $\begin{array}{c}
       0.8619\\
       0.5017 
     \end{array}$&
     $\begin{array}{c}
        194.882\\
        528.377
     \end{array}$&
     $\begin{array}{c}
        0.6282\\
        1.2213
     \end{array}$&
     $\begin{array}{c}
        0.1262\\
        0.1732
     \end{array}$&
     $\begin{array}{c}
        0.0\\
        180.0
     \end{array}$&
     $\begin{array}{c}
        162.492\\
        88.737
     \end{array}$&
     $\begin{array}{c}
        131.054\\
        207.200
     \end{array}$
\tabularnewline
\hline

     HD\thinspace160691 & $\begin{array}{c} b\\ c\end{array}$&
     $\begin{array}{c}
        1.5328\\
        1.1699
     \end{array}$&
     $\begin{array}{c}
        624.994\\
        2454.668
     \end{array}$&
     $\begin{array}{c}
        1.4684\\
        3.6550
     \end{array}$&
     $\begin{array}{c}
        0.3547\\
        0.4324
     \end{array}$&
     $\begin{array}{c}
        0.0\\
        180.0
     \end{array}$&
     $\begin{array}{c}
        76.468\\
        174.806
     \end{array}$&
     $\begin{array}{c}
        131.374\\
        178.448
     \end{array}$
\tabularnewline
\hline

     HD\thinspace202206 & $\begin{array}{c} b\\ c\end{array}$&
     $\begin{array}{c}
       17.4168 \\
       2.7195 
     \end{array}$&
     $\begin{array}{c}
        255.794\\
        1235.281
     \end{array}$&
     $\begin{array}{c}
        0.8302\\
        2.3623
     \end{array}$&
     $\begin{array}{c}
        0.4333\\
        0.4012
     \end{array}$&
     $\begin{array}{c}
        0.0\\
        180.0
     \end{array}$&
     $\begin{array}{c}
        161.125\\
        277.846
     \end{array}$&
     $\begin{array}{c}
        353.396\\
        71.407
     \end{array}$
\tabularnewline
\hline
     
   \end{tabular}
\vspace{3mm}
\caption{\label{tab_reduc}$V_0$, $rms$ and $\sqrt{\chi_\nu^2}$ values obtained
  for prograde and counter-revolving configurations. The values indicated for
  the prograde configurations come from~: (1) Vogt \etal (2005), (2) Lovis
  \etal (2006), (3) Tinney \etal (2006), (4) S\'andor \etal (2007), 
  (5) Cochran \etal (2007), (6) Butler \etal (2006), (7) Udry \etal (2002).}
   \begin{tabular}{ccccccccc}
\hline 
   \multicolumn{1}{c}{Systems} & 
   \multicolumn{1}{c}{ } & 
   \multicolumn{3}{c}{Counter-revolution configurations}& 
   \multicolumn{1}{c}{ } & 
   \multicolumn{3}{c}{Prograde configurations}\\
   \multicolumn{1}{c}{ } & 
            &$V_0$ $(m.s^{-1})$ & $rms$ $(m.s^{-1})$ & $\sqrt{\chi_\nu^2}$ & 
            &$rms$ $(m.s^{-1})$ & $\sqrt{\chi_\nu^2}$ & Ref.\\
\hline
\hline
    HD\thinspace37124& &$3.397$ & $5.008$  & $1.351$
                     & &$4.14-5.12$ & $0.96-1.14$ & (1)\\
\hline
    HD\thinspace69830& &$30289.729$& $0.808$  & $1.100$
                     & &$0.81$ & $1.095 $  & (2)\\
\hline
    HD\thinspace73526& &$-25.201$& $6.3398$    & $1.257$  
                     & &$8.04-8.36$ & $1.58-1.87$ & (3),(4) \\
\hline
    HD\thinspace108874& &$16.923$&$3.274$      & $0.386$
                      & &$3.7$   & $0.74$ & (1) \\
\hline
    HD\thinspace128311& &$-0.066$& $15.785$    & $1.785$
                      & &$18$    & $1.9$  & (1) \\
\hline
    HD\thinspace155358& &$10.751$& $5.904$  & $1.074$
                      & &$6.0$ & $1.15$ & (5) \\
\hline
    HD\thinspace160691& &$0.550$ & $3.469$     & $2.439$
                      & &$4.7$   & $1.1$ & (6)\\
\hline
    HD\thinspace202206& &$14706.445$& $8.517$  & $1.418$
                      & &$9.6$      & $1.5$ & (7)\\
\hline
   \end{tabular}
\end{center}
\end{table*}

\section{Method}

In this paper, we particularly focus on systems harboring planets with large
masses and close to their host star. As a consequence, the keplerian 
approximation is no longer suitable. It is necessary to perform dynamical fits
instead. 
We use a genetic algorithm (Pikaia, see Charbonneau 1995) with a set of
initial conditions  randomly taken in the orbital parameter space. 
We refer to Beaug\'e \etal (2007) for a complete description of the radial 
velocity method and the use of the Pikaia code.
Owing to the current theories of planetary formation (in a disk of gas and
dust) and to a large number of parameters to fit in the case of
orbital motions in a 3-dimensional space, dynamical fits are generally
performed while considering coplanar (and prograde) configurations. Hence,
the code we use was firstly developed for such prograde and coplanar
orbits. As a consequence, we have modified the Pikaia code in such a way that
observational fits can be performed for planetary systems harboring one planet
(whatever its location within the system)
in retrograde motion on its orbit (contrary to other planets of the same
system).

\clearpage

\begin{figure*}
\includegraphics[scale=0.45]{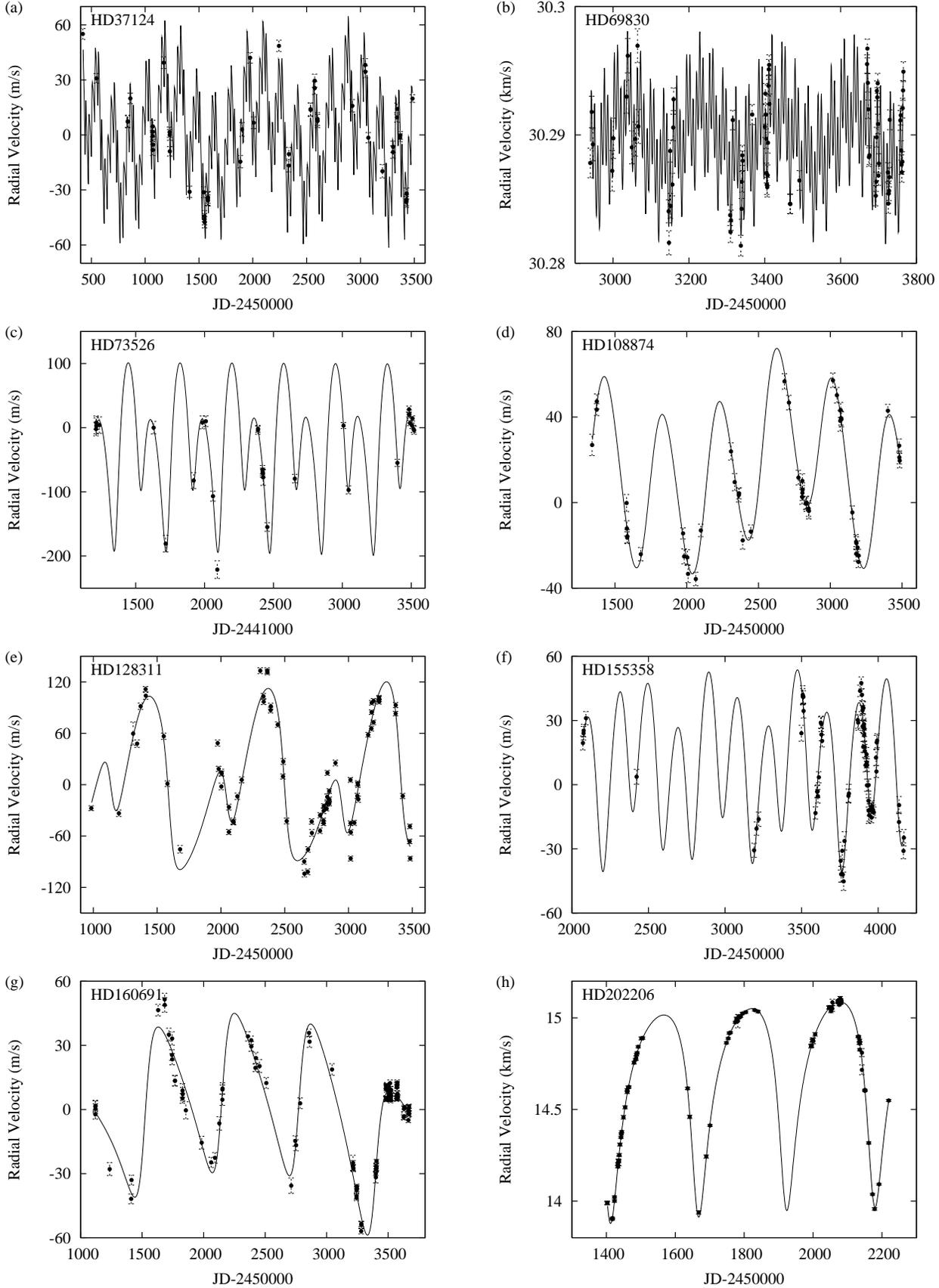}
   \caption{Dynamical fits with measured radial velocities of the
following planetary systems~: (a) HD\thinspace37124, (b) HD\thinspace69830,
(c) HD\thinspace73526, (d) HD\thinspace108874, (e) HD\thinspace128311, 
(f) HD\thinspace155358, (g) HD\thinspace160691 and (h) HD\thinspace202206. 
Orbital elements are presented in Table \ref{tab_obs}. Radial velocity
measurements are given in the references noted in Table \ref{tab_reduc}.}
\label{f1}
\end{figure*}
\clearpage

\section{Data reduction}
We have carried out dynamical fits for eight systems in counter-revolving
configurations (HD\thinspace37124, 
HD\thinspace69830, HD\thinspace73526, HD\thinspace108874, HD\thinspace128311, 
HD\thinspace155358, HD\thinspace160691, and HD\thinspace202206). In each case,
we have selected 
the best fits leading to dynamically stable solutions. While for most systems
a fit involving two planets is sufficient to obtain a rather good fit
(i.e. $\sqrt{\chi_\nu^2}$ close to $1$), the assumption of three planets for
the dynamical fits of the HD\thinspace37124 and HD\thinspace69830 systems is
necessary. The orbital elements found for the best fit of each system are 
presented in Table \ref{tab_obs} whereas the new values of 
$\sqrt{\chi_\nu^2}$ and $rms$ are compared, in Table \ref{tab_reduc}, to 
previous fits coming from prograde configurations (see Vogt \etal 2005, Lovis
\etal  
2006, Tinney \etal 2006, Butler \etal 2006, Cochran \etal 2007, S\'andor \etal 
2007, and Udry \etal 2002).
Some systems are henceforth found close to retrograde mean motion resonances 
(noted R-MMR hereafter)~: 
HD\thinspace73526 and HD\thinspace128311 (2:1 R-MMR), HD\thinspace108874 and
HD\thinspace160691 (4:1 R-MMR), HD\thinspace202206 (5:1 R-MMR),
HD\thinspace155358 (8:3 R-MMR).   
The new radial velocities are plotted in Fig. \ref{f1} according to the new 
dynamical fits of each studied planetary system. 

In most cases, $rms$ and $\sqrt{\chi_\nu^2}$ values obtained for
counter-revolving configurations are of the same order, and sometimes
slightly better, than for prograde configurations (see Table \ref{tab_reduc}).
Because the fit of the HD\thinspace160691 planetary system proves to be very 
bad ($\sqrt{\chi_\nu^2}=2.439$) while considering two planets in
counter-revolving configuration, we have tried to obtain a better result by
performing a new fit with  
three planets, one of them revolving in retrograde motion (not show here). 
However, we also find a high value of $\sqrt{\chi_\nu^2}$. The prograde fit
for this system proves to be definitely better.
In the end, for the other systems (except for the HD\thinspace37124 system),
the counter-revolving configurations are consistent with the current 
observational data.

\section{Discussion}
The dynamical fits presented in the present paper tend to show that, over the
eight studied multi-planetary systems, six of them are liable 
to be regulated by a mechanism involving a counter-revolving configuration
with a retrograde MMR. 
 Except for the HD\thinspace37214 and HD\thinspace160691 systems for
which the retrograde fits are indeniably bad, the whole of other 
fits are slightly better than fits in prograde configurations. 
Nevertheless, it remains necessary to perform new series of observations in 
order to enlarge the observational data samples and, as a consequence, to 
obtain more precise results. 

Although counter-revolving configurations seem possible both from an 
observational point of view (i.e. with observational consistence) and from a
theoretical one\footnote{See Gayon \& Bois (2008) and Gayon \etal (2009) for a
  theoretical study on the feasibility and efficiency of two planets to be in
  retrograde mean motion resonance.}, 
the formation of such systems does not seem obvious.
Indeed, the assumption that two giant planets are in a MMR and revolving
in opposite directions around their hosting star is apparently contradicting to
the most accepted formation theory of planetary systems, notably to the
formation and evolution of the resonant planetary systems (core accretion
mechanism combined by a planetary migration scenario). However, as mentionned 
in Gayon \& Bois (2008), two feasible processes leading to planets revolving in
opposite directions have been found. The first scenario has been introduced by 
Nagasawa \etal (2008). Starting from a
hierarchical 3-planet system and considering a migration mechanism including a
process of planet-planet scattering as well as a tidal circularization, the 
authors show that close-in planets may be  formed. In a few cases, due to the
Kozai mechanism, one planet may enter a retrograde motion. On the other hand, 
with Varvoglis, we have imagined a second feasible process that is related to
the capture of free-floating planets (private discussions). By integrating the
trajectories of planet-sized bodies that encounter a coplanar two-body system
(a Sun-like star and a Jupiter mass), Varvoglis has found that the probability
of capture is significant. Morever, the percentage of free-floating planets
forever captured is higher for retrograde motions than for prograde motions. 
As a consequence, it seems possible to find one day some planetary systems
stabilized in counter-revolving configurations. 

\section{Acknowledgments} 
Computations have been done on the ``M\'esocentre SIGAMM''
machine, hosted by Observatoire de la C\^ote d'Azur. Julie Gayon-Markt 
is supported through the NASA Postdoctoral Program.

{}

\end{document}